# Frequency-based image analysis of random patterns: an alternative way to classical stereocorrelation

J. Molimard[(a*)], G. Boyer[(b)], H. Zahouani[(c)], LTDS UMR CNRS 5513,

[a] molimard@emse.fr, University of Lyon, Laboratory of Tribology and Systems Dynamics UMR5513, École des Mines de Saint-Étienne, 158 Cours Fauriel, 42023 Saint-Etienne Cedex 2

[b] gaetan.boyer@peritesco.com, Laboratoire Peritesco, 7 Rue Sérane, 34000 Montpellier

c hassan.zahouani@ec-lyon.fr, University of Lyon, Laboratory of Tribology and Systems Dynamics UMR5513, Ecole Centrale de Lyon, 36 avenue Guy de Collongue, 69134 Lyon, France

## Abstract

The paper presents an alternative way to classical stereocorrelation. First, 2D image processing of random patterns is described. Sub-pixel displacements are determined using phase analysis. Then distortion evaluation is presented. The distortion is identified without any assumption on the lens model because of the use of a grid technique approach. Last, shape measurement and shape variation is caught by fringe projection. Analysis is based on two pin-hole assumptions for the video-projector and the camera. Then, fringe projection is coupled to in-plane displacement to give rise to 3D measurement set-up. Metrological characterization shows a resolution comparable to classical (stereo) correlation technique (1/100$^{th}$ pixel). Spatial resolution seems to be an advantage of the method, because of the use of temporal phase stepping (shape measurement, 1 pixel) and windowed Fourier transform (in plane displacements measurement, 9 pixels). Two examples are given. First one is the study of skin properties; second one is a study on leather fabric. In both cases, results are convincing, and have been exploited to give mechanical interpretation.

Keywords: 3D surface measurement, fringe projection, digital image correlation, soft tissue, skin

---

[*] Corresponding author

# 1. Introduction

Digital image correlation is one of the most diffused image processing technique among experimental mechanics community (Sutton, 1983). The system has been extended to different cases and firstly on warp surfaces (stereocorrelation) (Luo, 1993) (Garcia, 2002). But basically, one of the most interesting problems in DIC is the sub-pixel detection. Usually, authors use either correlation peak interpolation either sub-pixel deformation of the target sub-image. The deformation hypothesis can be either rigid body motion or a complete transformation function of the ZOI (Wattrisse, 2001). Recently, Réthoré (Réthoré 2008) proposed an original work, based on a correlation algorithm coupled with Finite Element based transformation functions, allowing for strongly regularized displacement fields with solid mechanics assumptions. In all these cases, the assumptions on the interpolation level are believed to have a direct influence on metrological properties of the correlation core (Bornert, 2009).

Beside the classical DIC approach, the grid method, even if less used, has rather comparable features: it is basically an image processing technique allowing for in-plane displacement fields. In the previous case, images are random patterns; in the latter, surface signature is a periodic grid. In this case, displacements derive from a spatial phase extraction (Surrel, 1994). The choice of surface encoding should orient the user toward one method or the other, but in fact, the nature of the pattern is not crucial: DIC has been successfully applied to periodic pattern for a very long time (Allais, 1994), and a random pattern can be seen has the superimposition of many frequencies. This last remark have been made recently by some authors (Poilâne, 2000) (Vanlanduit, 2009) (Molimard, 2008) and will be developed hereafter. Random pattern are of more practical use than periodic pattern for two basic reasons: first, periodic pattern generation is not as easy as someone could think, for very practical reasons. Basically anyway, surface preparation is easier for a random pattern than for a periodic one. Second, it is almost impossible to generate a periodic pattern on a non-flat surface – and develop a 3D surface grid method, even if industrial demand for measurement on real structures is high.

Finally, one should note that DIC is a genuine large-strain approach because it is based on re-correlation of decorrelated informations whereas Grid technique is a genuine small perturbation method because the phase difference is merging initial and final state of the investigated object, and because phase information is fairly more sensitive than amplitude. Consequently, it could be of great interest to adapt advances made in the context of grid techniques to random patterns, and to compare the results to those obtained by classical image correlation technique. We noted here the following specific "phase culture" items: camera distortion can be evaluated with a single grid image (Coudert, 2004), derivation can be derived analytically and the grid technique become sensitive to strain rather than displacements (Badulecu, 2009). Last, shape and shape variation can be detected in very good conditions using a video-projector and a camera (Sciammarella, 2005). This set-up gives high density information when classical stereovision is much more limited, as applications of fringe projection to surface rugosimetry outlines (Zahouani, 2009). This solution is slower than stereovision, but first applications of this technique will be under the context of biomechanics, and more precisely mechanics of skin, where speed is not crucial. Then, we consider that both informations of surface deformation (u,v,w) and surface topology are very complementary, and of equal interest.

## 2. 2D displacement method using frequency-based analysis

**2.1. DIGITAL IMAGE CORRELATION PRINCIPLE**

Assuming a reference image im0 described by $f(r,s)$, a deformed image im1 of im0 after a small strain is described by $g(r,s)$ by the following equation:

$$g(r,s) = f(r-\delta_x, s-\delta_y) + b(r,s) \qquad (1)$$

where $\delta x$ and $\delta y$ are the components of the displacement of im1 and $b(r,s)$ the noise measurement. A way to find $\delta x$ ans $\delta y$ is to maximize the function $h$ defined by:

$$h(r,s) = (g * f)(r,s) = \int_{-\infty}^{+\infty}\int_{-\infty}^{+\infty} g(a,b) f(a-r, b-s) da db \qquad (2)$$

where * denotes the cross-correlation product. Values of $r$ and $s$ found correspond to the maximal probability of displacement ($\delta x$, $\delta y$). This method can be applied in Fourier space using Fast Fourier Transform function, noted FFT2D. Equation 2 becomes:

$$g * h = FFT2D^{-1}\left(FFT2D(g)\overline{FFT2D}(f)\right) \qquad (3)$$

where the overline denotes the complex conjugate. Equation 2 and 3 are used at a local scale, on typical 32×32 zone of interest (ZOI). The work is repeated all over the map, giving a final displacement chart.

Now, classical digital image correlation performs the cross-product either in Fourier or real space. More refined approach exist by the way: for example, the signal could be normalized respect to the mean local intensity, and/or the mean local contrast (Bornert, 2009). The cross-correlation peak is commonly interpolated in order to reach a sub-pixel displacement accuracy. The interpolation function in not uniformly defined over the community. It could be for example a Gaussian or polynomial function. Basically, this choice doesn't have a strong theoretical basis, and authors have an empirical approach. We will propose hereafter an alternative of this weak point in the image correlation approach.

**2.2. SUB-PIXEL ALGORITHM**

The sub-pixel algorithm is based on the phase estimation of each ZOI. Because Fourier Transform requires continuity, the ZOI is windowed using a bi-triangular function. So far, the algorithm is an extension of the windowed Fourier Transform (WFT) algorithm proposed by Surrel (Surrel, 1997). As shown Fig. 1, in the frequency domain, each couple of frequencies is characterized by an amplitude and a phase. This phase is proportional to the displacement normal to the corresponding fringe direction. Note that the phase is defined only if the amplitude is higher than zero. Then, in absence of any phase jump, displacements can be related to any defined phase using the relationship:

$$\left\{\begin{array}{c}\vdots\\ \Delta\varphi_\theta^i\\ \vdots\end{array}\right\} = \left[\begin{array}{cc}\vdots & \vdots\\ \dfrac{2\pi}{p_\theta^i}\cos\theta^i & \dfrac{2\pi}{p_\theta^i}\sin\theta^i\\ \vdots & \vdots\end{array}\right]\left\{\begin{array}{c}\delta_x\\ \delta_y\end{array}\right\} = A\left\{\begin{array}{c}\delta_x\\ \delta_y\end{array}\right\} \qquad (4)$$

Displacements can then be derived from eq. 4 using the pseudo-inverse of A. This operation is possible if $\det\left(\left(A^t A\right)\right) \neq 0$. In practice, this means that at least two phases along two different directions exist.

$$\begin{Bmatrix} \delta_x \\ \delta_y \end{Bmatrix} = \left(A^t A\right)^{-1} A^t \begin{Bmatrix} \vdots \\ \Delta\varphi_\theta^i \\ \vdots \end{Bmatrix} \tag{5}$$

Practical problem of this approach is that the signal to noise ratio is weak for each couple of frequencies. Then, the quality of the measurement is obtained by averaging all the available informations throw the pseudo-inverse function. A detection of erroneous phases is implemented to increase the system resolution.

One should note also that a phase jump can occur in the Fourier domain. Even if a specific treatment should be developed, it sounds better to first use a pixel correlation algorithm. This ensures that the two ZOIs will be as superimposed as possible and that the fringe order is zero for any point and any couple of frequencies. No deformation of the ZOI is proposed here, considering that target applications will be in the small transformation domain.

## 2.3. CHARACTERIZATION

The method has been characterized using a simulation of a rigid body translation. A single ZOI is generated and translated from -0.5 to 0.5 pixel, in presence of noise or not. The ZOI has been over-sampled 10 times. 10 translation cases equally spaced are studied, and for each displacements 60 pairs of ZOI are sampled. Typical results are shown fig. 2 for a 32×32 region of interest, 12bit camera, with a 31 gray level white noise or not. One should note first that the error has the same amplitude as with image correlation techniques (Table 1), but the bias is fairly lower in this case. This should be one major advantage for this method.

Spatial resolution of the method is estimated using the following procedure: two independent noise distributions are added to the same speckle image. Then, displacements between the two situations are calculated, giving a displacement error map. The autocorrelation function of this error map is only affected by the displacement extraction procedure, so its size characterize the spatial resolution of the displacement extraction procedure. For the 32×32 window, effective spatial resolution is a diameter of 9 pixels (at 50% attenuation). This result should be surprising, but it can be partially explained by the use of the bi-triangular window: in 1D (the classical WFT algorithm) spatial resolution is half the window size. Then, it is worth noting that these values are very interesting compared with others zero-order image correlation algorithms.

**2.4. CAMERA DISTORTION**

Camera distortion is evaluated and corrected if necessary. The procedure consists in using a 2D regular grid (Coudert, 2004). A specific implementation of the Grid technique is used in order to evaluate accurately the grid step and to subtract the spatial carrier to the phase map. Differences between the phase map obtained experimentally and a flat phase map are due to the camera distortion (fig 3). Then, the phase map is converted to apparent displacements using the equation:

$$D_x = \frac{\Delta \varphi_x}{2\pi} P_{eff} \tag{6}$$

$$D_y = \frac{\Delta \varphi_y}{2\pi} P_{eff} \tag{7}$$

where $P_{eff}$ is the effective grid step in pixels.

Last, the images are corrected using the inverse deformation field $\{-\Delta D_x, -\Delta D_y\}^t$. Note that no camera model is necessary with this approach. A strong low-pass filter is recommended to avoid noise propagation.

# 3. 3D surface implementation

Coupling random speckle techniques (namely image correlation) to fringe projection is still an unuasual way to measure 3D displacement fields. Feasability has been demonstrated by Quan (Quan, 2004), and then used by Barrientos (Barrientos, 2008). Nam Nguyen (Nam Nguyen, 2009) added detection skills for non-continuous surfaces. We propose here to introduce a complete calibration of fringe projection technique in particular accounting for the variations of sensitivity due to videoprojector divergence, and the out-of-plane / apparent in-plane coupling due to camera divergence.

## 3.1. OPTICAL TEST-RIG

The optical set-up for 3D measurement is a classical fringe projection set-up, with a pocket-projector 3M MPRO 110, 800×600 and a CCD camera Imaging Source, 1280×960, 8bits. This solution is adapted to fields of investigation from 10×7 mm$^2$ to 200×150 mm$^2$ (see figure 4).

## 3.2. FRINGE PROJECTION

The fringe projection method has already been described by many authors (Sciammarella, 2005), (Peisen S. Huang, 2003), (Gigliotti, 2006), (Lagarde, 2002). The physical principle is straightforward: a periodic pattern of white and black lines is projected on an object; the light is diffused by the object and captured by a CCD video-camera. The deformation of the fringes, recorded as phase maps, has a known dependency to the out-of-plane displacements of the illuminated object.

The fringe projection technique exploits the light diffused by an object in order to measure its shape or shape variation, therefore the object must diffuse the light sufficiently. Moreover, in order to observe out-of-plane displacements, the angle between the projected fringes and the observed diffused light must not be null (fig. 5). Light intensities on an object illuminated by a set of fringes can be described by a periodic function $I_{li}$, with a perturbation $\varphi$ corresponding to the object shape:

$$I_{li}(x,y) = I_0(x,y)\left[1 + \gamma(x,y) \times \cos\left(\frac{2\pi}{p(x,y)}y + \varphi(x,y)\right)\right] \qquad (8)$$

This equation involves an average intensity $I_0$ and a contrast γ. These values should be constant over the whole map, but some low-frequency variations due to illumination inhomogeneities or diffusivity changes on top of the surface can occur. Consequently, both average intensity and contrast have to be considered as local quantities, typically calculated over few fringe periods, and can be denoted $I_0(x,y)$ and $\gamma(x,y)$. The pitch, $p$, is the distance between two light peaks on a flat surface. Again, due to perspective effects in particular, this pitch can change over the map, but this variation can be known either using a model or a calibration procedure. Last, the object is responsible for a phase shift $\varphi = \varphi(x,y)$ at each point of the field, as expressed by:

$$\varphi(x, y) = \frac{2\pi \times \tan\theta(x, y)}{p(x, y)} z(x, y) \qquad (9)$$

In this expression, the sensitivity characterized by the slope of the linear relationship between $\varphi(x,y)$ and $z(x,y)$, can be adjusted by modifying the pitch $p$ or the angle $\theta$ between the CCD video-camera and the video-projector. Again, it has to be noted that the sensitivity can vary locally. In particular, the video projector and the CCD camera commonly use divergent lens.

The classical pin-hole model is well adapted to such a configuration. Parameters of the model are:

- the camera magnification along the vertical axis ($\gamma_{CCD}$) and along the horizontal axis $\frac{\gamma_{CCD}}{\gamma_{CCD}}$,
- the distance between the CCD camera and the reference plane ($h_0$),
- the distance between the video-projector and the reference plane ($h_p$),
- the distance between the video-projector focal point and the CCD camera axis ($d$).

Measuring all these parameters is difficult in practice, and an inverse calibration is more adapted. Here, the calibration is based on the known rotation of a reference plane (Breque, 2004). The procedure is straightforward, but some hypothesis should be fulfilled: video-projector and camera axis should converge on a single point, this point being on the rotation axis; rotation axis should be perpendicular to the plane defined by the camera axis and the video-projector axis, and parallel to the fringes (fig. 4). Note that the system has to be calibrated after each geometrical changes in the configuration, but not before each new experiment.

Now, application of the pin-hole model gives the following set of equations:

$$z(r,s) = \frac{h_p h_o \left[ (2\pi f_p h_p - P_t d\varphi) \frac{\gamma_{CCD}}{\tau_{CCD}} \times r - P_t(d^2 + h_p^2) \varphi(r,s) \right]}{h_0 \left[ (2\pi f_p h_p - P_t d\varphi) d + P_t(d^2 + h_p^2) \varphi(r,s) \right] - h_p \left[ 2\pi f_p h_p - P_t d\varphi(r,s) \right] \frac{\gamma_{CCD}}{\tau_{CCD}} \times r}$$

$$x(r,s) = \frac{z(r,s) + h_0}{h_0} \frac{\gamma_{CCD}}{\tau_{CCD}} \times r \tag{10}$$

$$y(r,s) = \frac{z(r,s) + h_0}{h_0} \gamma_{CCD} \times s$$

The point $A(x, y, z)$ is known for any position in the object plane, referred by the co-ordinates $M(r, s)$. Note that $x$ and $y$ co-ordinates don't correspond to the $\left( \frac{\gamma_{CCD}}{\tau_{CCD}} \times r \;,\; \gamma_{CCD} \times s \right)^t$ because of the perspective effect on the camera.

Extraction of the phase from intensity map(s) requires either spatial or temporal phase shifting techniques. The Photomecanix toolbox, developed in the laboratory, has genuine implementation of both techniques, as prescribed by Surrel (Surrel, 1997). The choice only depends on the situation: if temporal effects are expected, spatial phase shifting is more appropriate, because it only requires one image (Wang, 2010). If not, temporal phase shifting technique should be preferred for its higher spatial resolution. Only this method is briefly described here.

A set of $n \times q$ fringe patterns with a known phase shift $q/2\pi$ is projected successively on the surface, first and last fringe pattern being shifted by a $n \times 2\pi$, $n \in \mathbb{Z}$ phase. Then, the intensity variation at each point (i.e. each camera pixel) corresponds to a sine wave function with an initial phase shift. The phase is evaluated using the Fourier Transform:

$$\varphi(r,s) = \arctan_{2\pi} \left( \frac{\sum_{k=1}^{nq} \left\{ \sin\left( k \cdot \frac{2\pi}{q} \right) I_k(r,s) \right\}}{\sum_{k=1}^{nq} \left\{ \cos\left( k \cdot \frac{2\pi}{q} \right) I_k(r,s) \right\}} \right) \tag{11}$$

Metrological performances of the shape measurement set-up are interesting compared to the classical stereovision technique: the spatial resolution is 1 pixel (8 to 156 µm, depending on the field of view), and the resolution is $\sigma = 1/100^{th}$ fringe, i.e. 10 µm at best. This capacity is very important for high frequency phenomena monitoring: skin submitted to mechanical load, metal instability under forming process, ...

### 3.3. COUPLING OUT-OF-PLANE AND IN-PLANE DATA

Up to the author's knowledge, the coupling of in plane and out-of-plane displacements in such a context hasn't been described yet. In fact, one should consider the following problems:

- in-plane displacements are the apparent displacements on the image plane. These apparent displacements are affected by the shape of the object if the CCD lens is not telecentric.

- When the object is deformed in plane, the out-of-plane displacement is no longer the difference of the two shapes. Physical points should be clearly identified and subtracted one to the other.

Finally, displacements are calculated as the difference between *x*-, *y*- and *z*- co-ordinates before and after deformation. Figure 6 presents a surface submitted to in-plane and out-of-plane displacements. The prime symbol denotes deformed state ; ($\delta x$, $\delta y$) is the apparent displacement at the image plane. Initial shape is considered as the reference shape, so deformed shape, known at point (*r,s*) is interpolated at (*r+δx, s+δy*), so:

$$u = x'(r+\delta x, s+\delta y) - x(r,s)$$
$$v = y'(r+\delta x, s+\delta y) - y(r,s) \quad (12)$$
$$w = z'(r+\delta x, s+\delta y) - z(r,s)$$

Finally, last step is the projection of the given displacement on the local co-ordinate system. This operation implies knowing the first derivative of $z(x, y)$. This operation has the main disadvantage of propagating the noise. So far, a low-pass filter is necessary, and shape discontinuity should affect projected displacement fields.

## 4. Examples

**4.1. SIMULATED EXPERIMENT**

First test on the sub-pixel algorithm has been conducted on a fake displacement map provided by GDR 2519 (Bornert, 2009). It consists in a sine-wave displacement field, as shown figure 7 encoded on a 8-bits intensity map, without noise. Here, the displacement amplitude has been set to 0.02 pixels. Root mean square error found both in $x$ and $y$ directions is 0.003 pixels.

**4.2. SKIN PROPERTIES EVALUATION**

Skin properties evaluation is a major topic for cosmetic industry as well as for reconstructive surgery. Many tests have been proposed in the literature (suction test, indentation test, ...). Here, we used an in-plane tension test, corresponding to a classical surgeon handy test. The objective of this work is to identify the elastic anisotropic properties. All this mechanical development is out of purpose here, and we will focus on the measurement only. Readers should refer to (Boyer, 2009) for further details.

The surface signature is directly the skin furrows, and no synthetic pattern is necessary. Unfortunately anyway, during the loading process, furrows are changing and the optical flow conditions are no longer valid, leading to erroneous points. Besides, mean strain level is up to 10%. The retained solution consists in measuring small gap displacements (and consequently without any changes in the surface signature), and summing the intermediate states. Because the 1$^{st}$ and last step correspond to different geometrical configurations, the summation has to be performed in the initial state. Procedure developed by Avril (Avril 2008) is used: each displacement map is interpolated by a finite element mesh. Then, these maps are continuous, and it is possible to interpolate analytically mesh points for one load step to the other.

Figure 8 shows displacement fields obtained on the same specimen (a 26 year old young man), with different orientations of the loading/measuring device with the arm. The covered field is 10×13 mm$^2$, and the final displacement is 1 mm along the vertical (*y*) axis. The 3 set of results present different kind of asymmetry with the vertical axis. Because the quality of the loading system has been verified, this asymmetry clearly indicates the effect of the orientation of the loading apparatus on the displacement maps, and consequently the apparent anisotropy of the skin.

Note that the evaluation procedure has to be strictly non invasive for ethic reasons. The proposed solution, based on an optical technique, using the natural surface signature, completes perfectly this requirement. Last, load step has to be low enough not to damage the skin.

### 4.3. LEATHER FABRIC UNDER TENSION

Second example proposed is the study of leather fabric under tension. We focused here on the mechanical linkage between the fabric and a screw. The fabric is folded at the beginning of the test, so 3D displacements are expected. Figure 9 shows the shape of the leather fabric before loading, displacement vectors, loading direction and the global frame of reference. Here, only one quarter of the surface is represented. The leather is clamped with two screws on the movable jaw and two screws on the fixed jaw. The 3D representation shows the dominant out-of-plane movement, and a global in-plane rotation.

Displacements in the global or local frame of reference are presented figure 10. An additional high frequency noise is visible if using the local frame of reference. Anyway the signal to noise ratio is good enough to exploit the data. Note that the results are given without any low-pass filter. Few changes can be seen for the out-of-plane movement, but clear differences appear for the two in-plane displacements (U and $U_t$, V and $U_t'$). In particular, flexure effect along direction y is obvious if using the local frame of reference (see $U_t$).

# 5. Conclusion and perspectives

3D measurements of shape, shape variation, and displacements on non-flat surfaces are of critical practical interest, either in mechanical engineering and biomechanics. This paper presents an original way to measure 3D shape and deformations with geometrical techniques. In-plane deformation is accurately measured by frequency-based image analysis of random patterns, and out-of-plane component is caught using temporal phase-stepping fringe projection. The proposed set-up potentially takes advantage of all the developments in phase processing. The link between in-plane and out-of-plane measurements, based on the pine-hole model has been clearly exposed.

First results show very promising performances. In particular, the main advantage is the very good final spatial resolution, and a low bias: shape resolution is 1 pixel; displacements resolution is 8 pixels. The in-plane and out-of-plane resolutions are few hundredth pixels.

Two examples illustrate the first applications of the technique: skin behaviour under tension on a living body, and leather fabric under tension. In both cases, signal to noise ratio is high enough, and post-processing is not necessary. Complete uncertainty analysis will be developed in the future, anyway we consider the system mature enough to serve as a reference tool for biomechanical studies.

Acknowledgements:

Authors would like to thank Dr. Avril for his help to this work.

References:

Allais L., Bornert M., Bretheau T., Caldemaison D., (1994) Experimental characterization of the local strain field in a heterogeneous elastoplastic material, Acta Metallurgica et Materialia, 42 (11), 3865-3880.

Avril S., Feissel P., Pierron F., Villon P., (2008) Estimation of the strain field from full-field displacement noisy data. Comparing finite elements global least squares and polynomial diffuse approximation, European Journal of Computational Mechanics, Vol 17 (5-7), 857-868

Badulescu C., Grédiac M., Mathias J.D., Roux D., (2009) A Procedure for Accurate One-Dimensional Strain Measurement Using the Grid Method, 49 (6), 841-854.

Barrientos B., Cerca M., Garcıa-Marquez J., Hernandez-Bernal C., (2008) Three-dimensional displacement fields measured in a deforming granular-media surface by combined fringe projection and speckle photography, Journal of Optics A: Pure and Applied Optics 10, 104027 (10pp).

Bornert M., Brémand F., Doumalin P., Dupré J.C., Fazzini M., Grédiac M., Hild F., Mistou S., Molimard J., Orteu J.J., Robert L., Surrel Y., Vacher P., Wattrisse B., (2009) Assessment of Digital Image Correlation measurement errors: Methodology and results, Experimental Mechanics 49 (3), 353-370.

Boyer G., Molimard J., Zahouani H., Pericoi M., Avril S., (2009) Assessment of biomechanicals properties of human skin by a new device combining mechanicals and optical measurements, Proceedings of the International Conference on Bioengineering & Biomaterials, 18-20 March 2009, Meknes, Morocco, 10pp.

Breque C., Dupre J.C., Bremand F., (2004) Calibration of a system of projection moiré for relief measuring: biomechanical applications, Optics and Lasers in Engineering, 41 (2), 241-260.

Coudert S., Triconnet K., Zebiri A., Surrel Y., (2004) Étalonnage transverse d'un objectif de caméra par la méthode de la grille, colloque Photomécanique, Albi, France, 4-6 May 2004. (in French).

Garcia D., Orteu J.J, Penazzi L, (2002) A combined temporal tracking and stereo-correlation technique for accurate measurement of 3D displacements: application to sheet metal forming, Journal of Materials Processing Technology, 125-126, 736-742.

Gigliotti M., Molimard J., Jacquemin F., Vautrin A., (2006) On the nonlinear deformations of thin unsymmetric 0/90 composite plates under hygrothermal loads, Composites Part A: Applied Science and Manufacturing, 37(4), 624-629.

Lagarde J.M., Rouvrais C., Black D., Diridollou S., Gall Y., (2002) Skin topography measurement by interference fringe projection: a technical validation, 7 (2), 112 – 121.

Luo, P.F., Chao Y.J., Sutton M.A., Peters III, W.H., (1993) Accurate measurement of three-dimensional deformations in deformable and rigid bodies using computer vision, Experimental Mechanics, 33 (2), 123-132.

Molimard J., (2008) Frequency-based image analysis of random patterns: a first attempt, Proceeding of Photomechanics'08, Loughborough, UK, 7-9 July 2008.

Nam Nguyen T., M Huntley J., Burguete R., Russell Coggrave C., (2009) Combining digital image correlation and projected fringe techniques on a multi-camera multi-projector platform, Journal of

Physics: Conference Series, 181, 012076 (8pp).

Peisen S. Huang, Chengping Zhang, and Fu-Pen Chiang, (2003) High-speed 3-D shape measurement based on digital fringe projection, Optical Engineering, 42, 163-168.

Poilâne C., Lantz E., Tribillon G., Delobelle P., (2000) Measurement of in-plane displacement fields by a spectral phase algorithm applied to numerical speckle photography for microtensile test, European Physical Journal: Applied Physics, 11, 131-145.

Quan C., Tay C.J., Huang Y.H., (2004) 3-D deformation measurement using fringe projection and digital image correlation, Optik 115 (4), 164-168.

Réthoré J., Hild F., Roux S., (2008) Extended digital image correlation with crack shape optimization, International Journal of Numerical Methods in Engineering, 73 (2), 248-272.

Sciammarella C.A., Lamberti L., Sciammarella F.M., (2005) High-accuracy contouring using projection moiré, Optical Engineering, 44 (9), 093605 (12pp.).

Surrel Y. (1994) Moiré and grid methods in optics, *SPIE 2342*, 213-220.

Surrel Y., (1997) Additive noise effect in digital phase detection, Applied Optics, 36 (1), 271-276.

Sutton M.A., Wolters W.J., Peters W.H., Ranson W.F., McNeill S.R., (1983) Determination of displacements using an improved digital image correlation method, Image and Vision Computing, 1 (3), 133-139.

Vanlanduit S., Vanherzeele J., Longo R., Guillaume P., (2009) A digital image correlation method for fatigue test experiments, Optics and Lasers in Engineering, 47, 371– 378.

Wang P., Drapier S., Molimard J., Vautrin A., Minni J.C., (2010) Characterization of Liquid Resin Infusion (LRI) filling by fringe pattern projection and in situ thermocouples, Composites Part A: Applied Science and Manufacturing, 41(1), 36-44.

Wattrisse B, Chrysochoos A, Muracciole J-M, Némoz-Gaillard M (2001) Analysis of strain localization during tensile tests by digital image correlation. Exp Mech 41(1), 29–39.

Zahouani H., Vargiolu R., Skin morphology and volume: methods of evaluation, in Injection treatments in cosmetic surgery, Ascher B. and al. ed., Informa, London 2009, 20 p.

Table caption

Table 1. Error quantification for a 32×32 window.

figure captions

Figure 1. Basic principle of sub-pixel algorithm.

Figure 2. Quality of identified displacements for a pure translation without and with noise.

Figure 3. Distorsion of an Optem Zoom 125.

Figure 4. Optical set-up and calibration test-rig.

Figure 5. Fringe projection basic principle.

Figure 6. Coupling in-plane and out-of-plane displacements.

Figure 7. in-plane displacement maps for a simulated experiment.

Figure 8. Intensity maps on a skin study before and after loading.

Figure 9. Initial shape and displacement vectors.

Figure 10. Displacement fields in the global frame of reference (left) and projected on the surface (right).

|  | 0 GL noise | 31 GL noise |
| --- | --- | --- |
| Resolution (in pixels) | 0.025 | 0.035 |
| Bias (in pixels) | 0.0044 | 0.0056 |

**Table 1. Error quantification for a 32×32 window.**

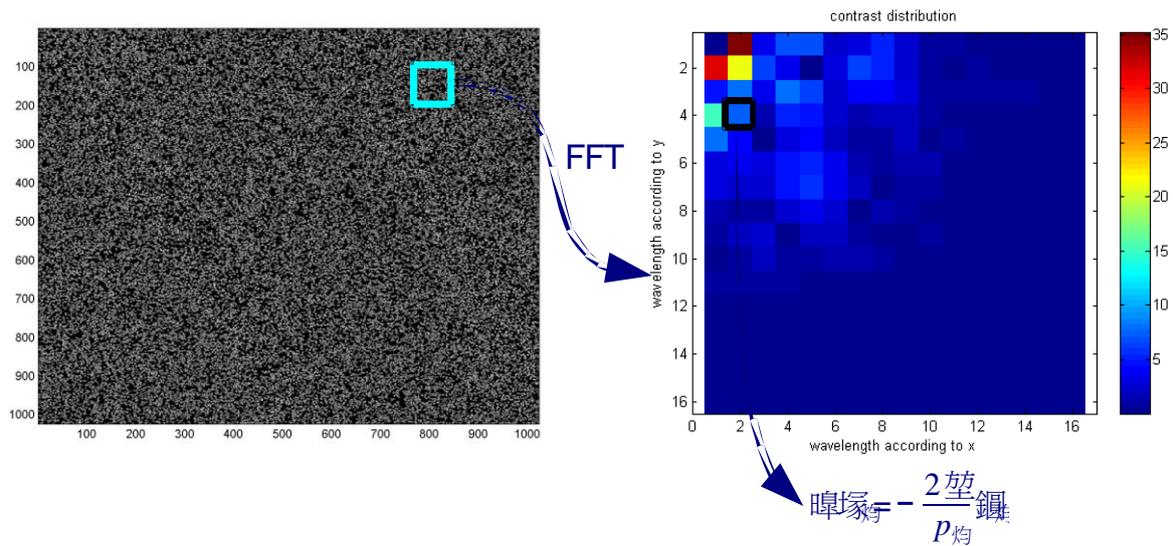

**Figure 1. Basic principle of sub-pixel algorithm.**

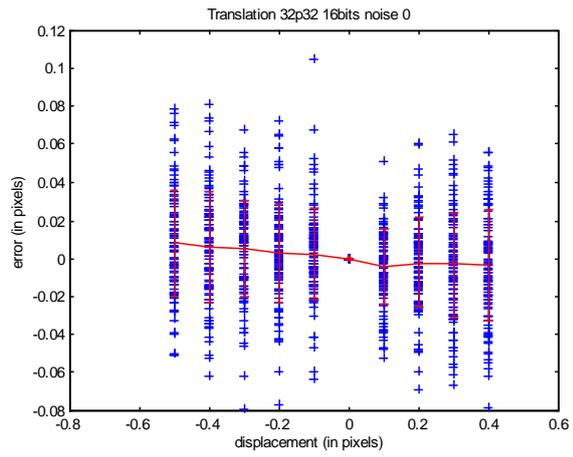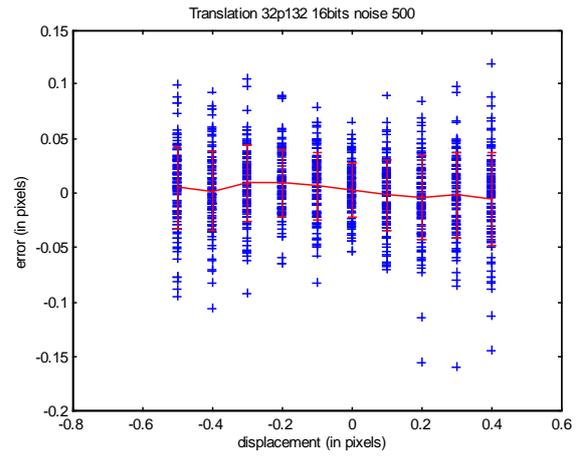

**Figure 2. Quality of identified displacements for a pure translation without and with noise.**

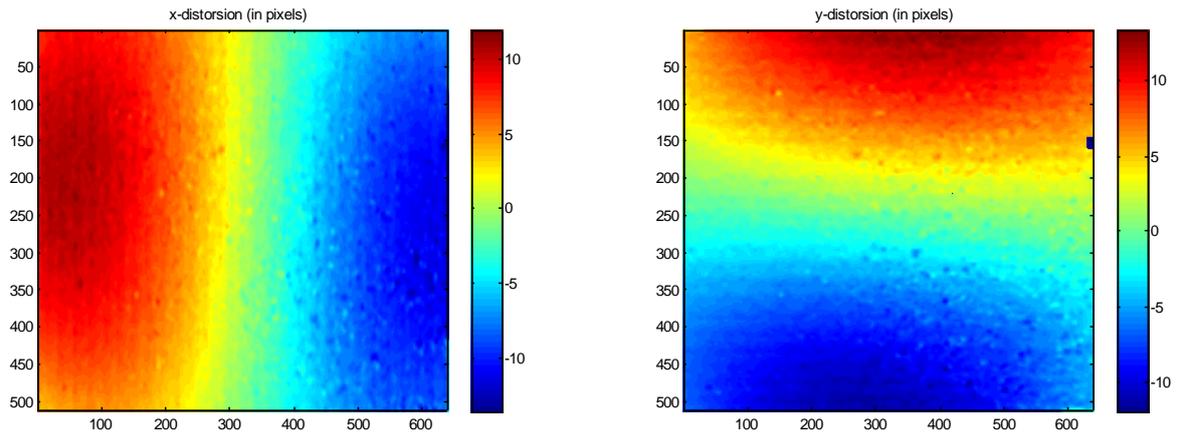

**Figure 3. Distorsion of an Optem Zoom 125.**

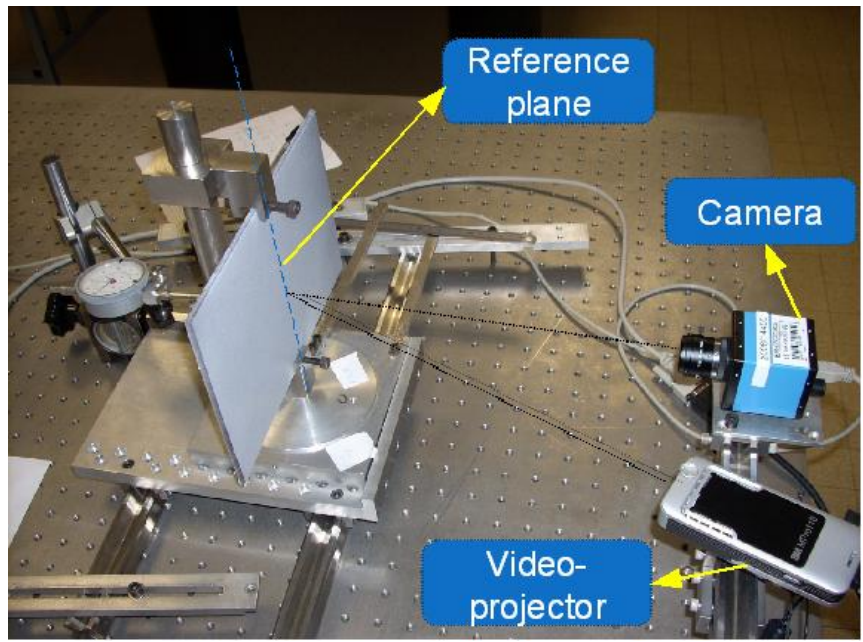

**Figure 4. Optical set-up and calibration test-rig.**

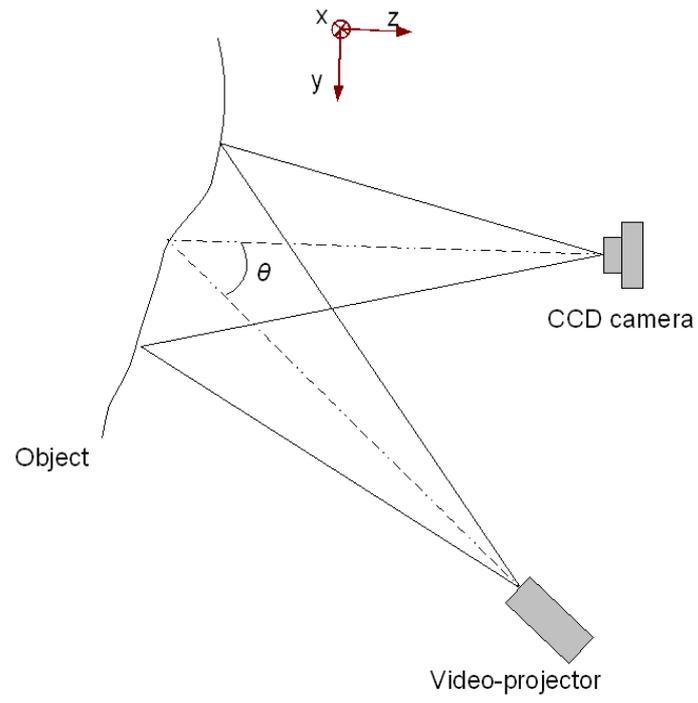

**Figure 5. Fringe projection basic principle.**

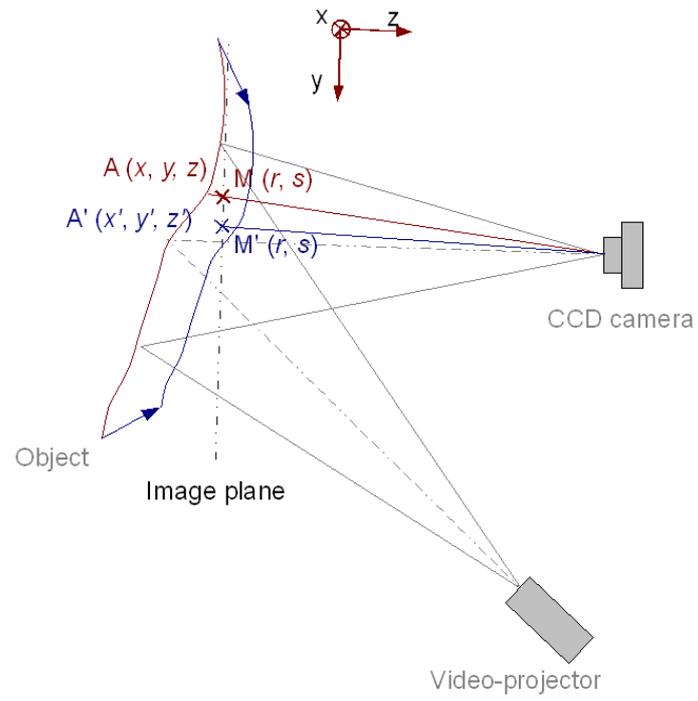

**Figure 6. Coupling in-plane and out-of-plane displacements.**

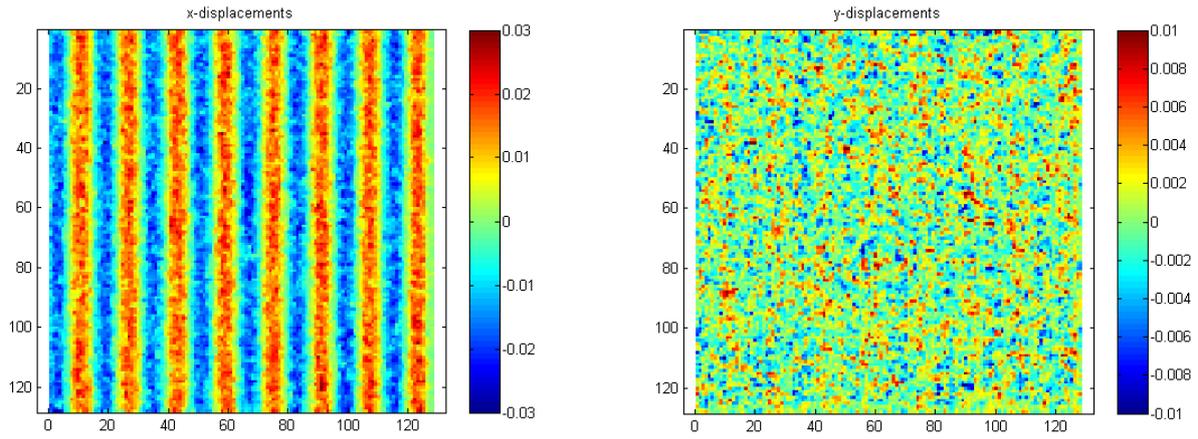

**Figure 7. in-plane displacement maps for a simulated experiment.**

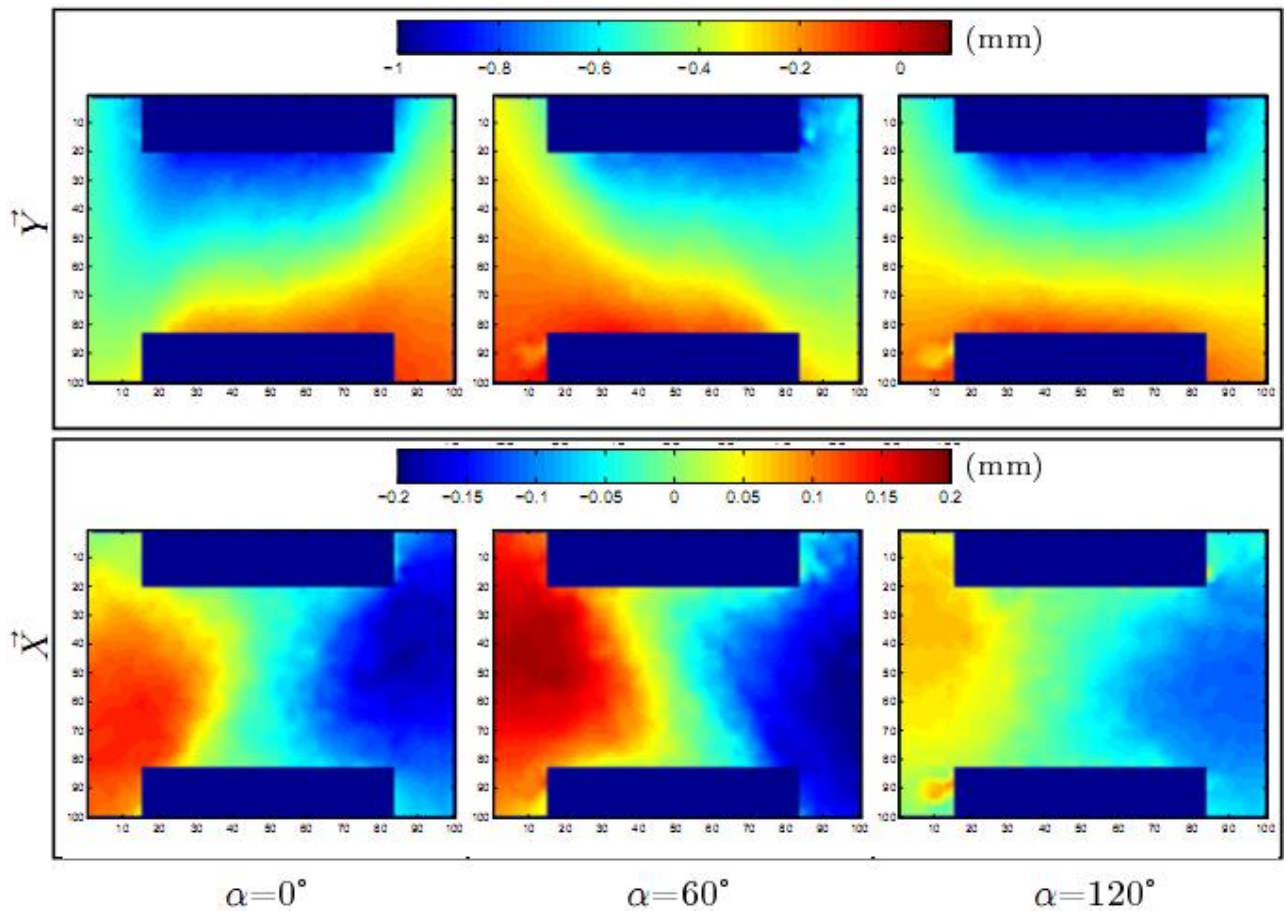

**Figure 8. Intensity maps on a skin study before and after loading.**

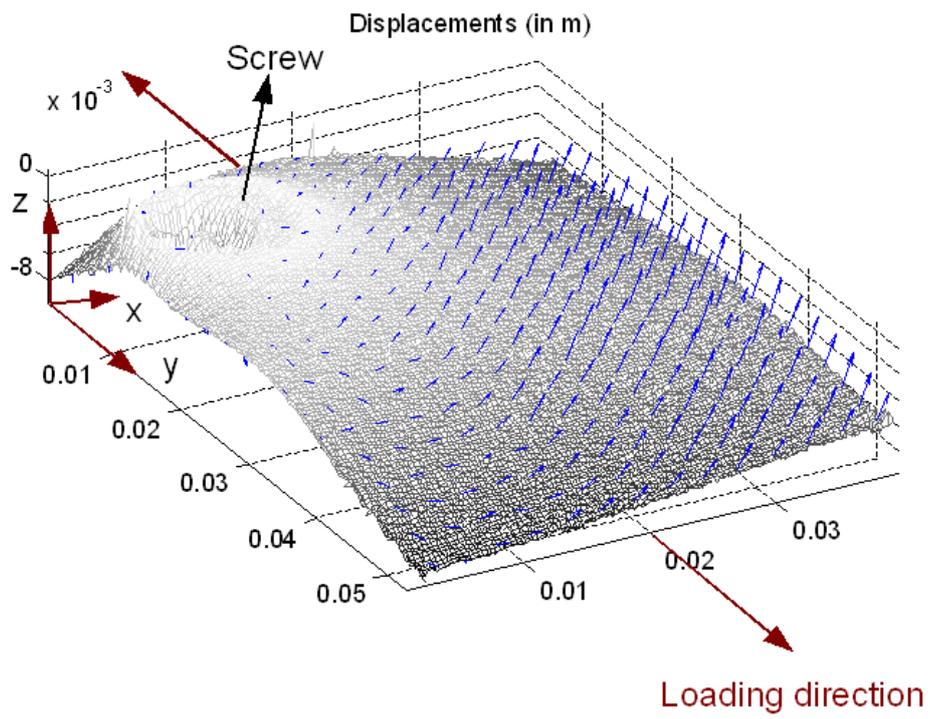

**Figure 9. Initial shape and displacement vectors.**

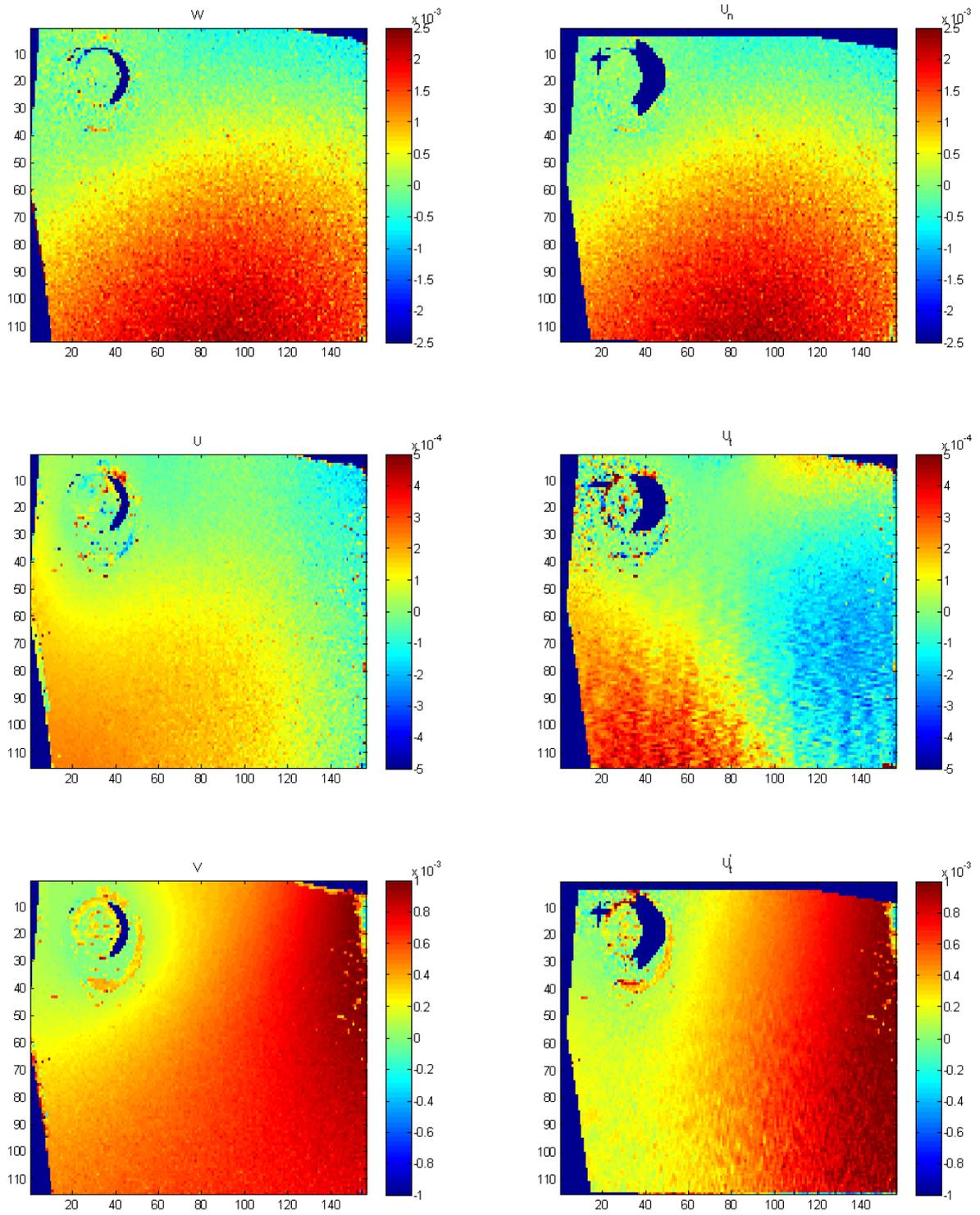

**Figure 10. Displacement fields in the global frame of reference (left) and projected on the surface (right).**